# MSSM Based Theory for Planck Precision Results (2018), Dark Matter, Dark Energy and Solution of Cosmological Problems

Debatosh Majumdar*

**Abstract**

This paper shows that the precision measurement data of the cosmological parameters obtained by the Planck Satellite Observatory (2018) from the Cosmic Microwave Background Anisotropy, can be calculated with excellent agreement with these measured values by considering the masses of the Gauge Bosons and their respective supermultiplet partners from the combination equations of the decay products of these gauge supermultiplets of the MSSM theory. Therefore, these results support Einstein's Lambda-CDM (ΛCDM) cosmological standard model. Further, these equations are useful to calculate the masses of the ordinary nucleon, proposed anti-supersymmetric nucleon (considered as the dark matter of the universe in place of LSPs), and Higgs boson in agreement with their respective experimental masses or density ratios. In addition, this theory solves the problems of baryon number violation, lepton number violation, and matter-antimatter asymmetry of the universe. It is expected that the masses of the winos and anti-supersymmetric nucleons considered here will be verified from the decay of the Higgs boson production in the interaction of top-anti-top quarks in the latest high energy hadron collider experiments.

* Ex-Principal,
Acharya Jagadish Chandra Bose College,
University of Calcutta,
1/1A, A. J. C. Bose Road, Kolkata, India.

email: debatosh.majumdar@gmail.com

# 1 Introduction

The motivation of this paper is to present an appropriate particle theory for the Planck precision measurement results 2018 of the anisotropy parameters of the Cosmic Microwave Background (CMB) by the Planck Collaboration (2020) and to use this theory to explain the dark matter (DM), dark energy (DE), and some other problems of the universe.

Measurements of most of the cosmological parameters from the observed CMB anisotropies were made by the Wilkinson Microwave Anisotropy Probe (WMAP) Collaboration during the period (2001-2009) [1]. This was followed by the Planck satellite observatory collaboration which published their results from time to time. Subsequently, the Particle Data Group (2019) reviewed all these astrophysical parameter values and published them in PDF format [2]. Finally, the Planck collaboration made more precision measurement of those parameters in 2018 and published them in 2020 [3]. These results were based on the Einstein's Lambda-CDM (ΛCDM) cosmological standard model (CSM). However, the problem of this and other cosmological models is that none of them can theoretically determine the values of these parameters to compare them with their respective observed values. Recently, there was an attempt by J. S. Farnes for a unifying theory of dark energy and dark matter based on the hypothesis of negative-positive masses within a modified ΛCDM framework [4]. But his theory shows only the expressions of the dark matter and dark energy masses in terms of these masses, but no parameter values of them; and further, no negative mass particle has yet been observed experimentally. More recently, there was another approach by M. J. Longo to explain both DM and DE by the existence of long-lived topological vortices. But this theory also cannot calculate their theoretical parameter values as the vortices mass is not known [5].

It is known that the Standard Model (SM) theory of Big Bang Nucleosynthesis (BBN) can account for the cosmologically observed baryon density of the universe [6-9]. But for DM density, the SM has no theory at all. Under the circumstances, there are many theories beyond the SM for the DM particles. In contrast, there are people who would reject the existence of DM and replace the Lambda-CDM modal by a non-relativistic MOND or MOND-like relativistic theories to explain cosmological findings. But so far none of these theories is established [10-22].

As regards DE, it was first supposed to be the same as the Einstein cosmological constant. But this approach of SM leads to a value ~ 60 orders of magnitude larger than the accepted value of this constant [23]. So, it is now described as an unknown kind of negative vacuum energy of constant density causing the accelerated expansion of the universe [24-27].

The minimal supersymmetric standard model (MSSM) introduced as an extension of the SM was primarily intended to solve the DM problem [28-35]. It did successfully solve the unification of the forces (GUT) at an energy scale of $10^{16}$ $GeV$ and predicted proton decay [36-39] though this decay was not observed in any LHC or XONON experiment, e.g., ref. [40]. In the MSSM the lightest supersymmetric particles (LSPs), are the preferred candidates for the DM particles. There are also theories where either heavy neutrinos or axions are considered as the DM particles. But till now there is no experimental evidence for any of these particles either from LHC or XENON nT [41-48].

In the above-mentioned theories, DM particles are non-baryonic and so cannot solve the baryon anisotropy problems of the universe. However, in this paper my motivation is to use the basic properties of MSSM to make a theory for the determination of Planck precision cosmological parameter values and baryonic DM particles to solve the anisotropy problems. In section 2, the necessary theories for this purpose are mentioned and in section 3, the resulting equations are described, and used to calculate the parameter values of the Planck precision results. In section 4, solutions for the problems of baryon- and lepton- numbers violation, and matter-antimatter asymmetry are discussed. In section 5, the paper is concluded by summarizing the results obtained and mentioning their prospects.

## 2 The theory for the parameter values of the lambda CDM model

Let us consider the MSSM theory during the epoch of cosmic inflation after the Big Bang when the gauge bosons and their respective superpartners (gaginos) were in the plasma states along with their interacting virtual particles of all the forces. It is obvious that as the space was expanding and the temperature dropped, the interacting particles recombined and formed natural nucleons and anti-supersymmetric nucleons along with leptons, anti-supersymmetric leptons and gauge bosons of the universe. The total process is explained by the application of MSSM for theoretical determination of the Planck observational cosmological parameter results as described below.

The MSSM theory is based on the minimal chiral and vector supermultiplets as shown in the flowing Table 1 and Table 2 respectively.

TABLE 1

| Names | Symbol | Spin 0 | | Spin ½ | |
|---|---|---|---|---|---|
| squarks, quarks | $Q_L$ | $\tilde{u}_L^{2/3}$ | $\tilde{d}_L^{-1/3}$ | $u_L^{2/3}$ | $d_L^{-1/3}$ |
| anti-squarks, anti-quarks | $\overline{Q_R}$ | $\tilde{u}_R^*$ | $\tilde{d}_R^*$ | $u_R^\dagger$ | $d_R^\dagger$ |
| sleptons, leptons | $L_L$ | $\tilde{\nu}_L^0$ | $\tilde{e}_L^{-1}$ | $\nu_L^0$ | $e_L^{-1}$ |
| anti-slepton, anti-lepton | $\overline{e_R}$ | - | $\tilde{e}_R^*$ | - | $e_R^\dagger$ |
| Higgs, higgsinos | $H_u$ | $H_u^+$ | $H_u^0$ | $\widetilde{H}_u^+$ | $\widetilde{H}_u^0$ |
| | $H_d$ | $H_d^0$ | $H_d^-$ | $\widetilde{H}_d^0$ | $\widetilde{H}_d^-$ |

Table 1. Chiral supermultiplets of the MSSM. The spin-0 fields are complex scalars, and spin-½ fields are left-handed two component Weyl fermions. Their spin 0 antiparticle is the right-handed conjugate particle while the spin-half antiparticle is the right-handed adjoint particle. In this field structure, each spin-0 squark and anti-squark has the same mass, charge, baryon number and color as those of its spin half superpartner. The lepton superpartners do not carry colors and instead of baryon numbers, they carry equal lepton numbers along with their equal charge and mass. Further, they do not have anti-neutrino superpartners. Along with these supermultiplets the composite Higgs and their corresponding superpartners are also shown here.

TABLE 2

| spins | | 1. spin ½ | 2. spin 1 |
|---|---|---|---|
| Names | | Gauginos | Gauge Bosons |
| 1. gluinos, | 2. Gluons | $\tilde{g}$ | $g$ |
| 1. winos, zeno | 2. W-, Z- bosons | $\widetilde{W}^{\mp}$ $\widetilde{W}^{0}$ $\tilde{Z}^{0}$ | $W^{\pm}$ $W^{o}$ $Z^{o}$ |
| 1. bino, | 2. B-boson | $\tilde{B}^{0}$ | $B^{o}$ |

Table 2. Vector gauge supermultiplets of the MSSM. Both gluons and gluinos are $\mathrm{SU}(3)$ color-anticolor octets together with their colorless $\mathrm{U}(1)$ singlets of strong forces. Similarly, winos and the zeno together with W- and Z- bosons carry the electroweak forces under $\mathrm{SU}(2)$ and $\mathrm{U}(1)$ symmetry while the bino and the B-boson carry the EM force under $\mathrm{U}(1)$ symmetry.

Now, in SM vacuum polarization we consider a virtual pair production and annihilation of a spin 1/2 fermion and its anti-partner by a spin 1 gauge boson as shown in Table 3 [49: see ch. 2, pp. 66-78]. Similarly, in MSSM we consider the spin 1/2 gaugino, too, which is the superpartner of the gauge boson (Table 2) to decay into a particle and an anti-superparticle as shown in the same Table 3.

TABLE 3

| spin ½ gaugino decay | spin 1 Boson decay |
|---|---|
| $\tilde{g} \rightarrow u^{2/3} + \tilde{u}^{*-2/3}$ | $g \rightarrow u^{2/3} + u^{\dagger-2/3}$ |
| $\tilde{g} \rightarrow d^{-1/3} + \tilde{d}^{*+1/3}$ | $g \rightarrow d^{-1/3} + d^{\dagger 1/3}$ |
| $\widetilde{W}^{+} \rightarrow u^{2/3} + \tilde{d}^{*1/3}$ | $W^{+} \rightarrow u^{2/3} + d^{\dagger 1/3}$ |
| $\widetilde{W}^{+} \rightarrow \nu^{0} + \tilde{e}^{*+1}$ | $W^{+} \rightarrow \nu^{0} + e^{\dagger+1}$ |
| $\widetilde{W}^{-} \rightarrow d^{-1/3} + \tilde{u}^{*-2/3}$ | $W^{-} \rightarrow d^{-1/3} + u^{\dagger-2/3}$ |
| $\widetilde{W}^{-} \rightarrow e^{-1} + \tilde{\nu}^{*0}$ | $W^{-} \rightarrow e^{-1} + \nu^{\dagger 0}$ |
| $\widetilde{W}^{0}, \tilde{Z}^{0} \rightarrow u^{2/3} + \tilde{u}^{*-2/3}$ | $W^{0}, Z^{0} \rightarrow u^{2/3} + u^{\dagger-2/3}$ |
| $\widetilde{W}^{0}, \tilde{Z}^{0} \rightarrow d^{-1/3} + \tilde{d}^{*+1/3}$ | $W^{0}, Z^{0} \rightarrow d^{-1/3} + d^{\dagger 1/3}$ |
| $\widetilde{W}^{0}, \tilde{Z}^{0} \rightarrow \nu^{0} + \tilde{\nu}^{*0}$ | $W^{0}, Z^{0} \rightarrow \nu^{0} + \nu^{\dagger 0}$ |
| $\widetilde{W}^{0}, \tilde{Z}^{0} \rightarrow e^{-1} + \tilde{e}^{*+1}$ | $W^{0}, Z^{0} \rightarrow e^{-1} + e^{\dagger+1}$ |
| $\tilde{B}^{0} \rightarrow e^{-1} + \tilde{e}^{*+1}$ | $B^{o} \rightarrow e^{-1} + e^{\dagger+1}$ |

Table 3. Decay products of the gauge bosons and the gauge fermions of the gauge supermultiplets. For simplicity of writing, I have dropped here the chirality indices (L, R), color indices (r, b, g) and other quantum numbers of a particle but used the opposite sign for the charge of the particle along with a star sign (*) to denote the chiral supersymmetric antiparticle, and an adjoint sign (†) to denote the chiral SM antiparticle as shown in Table 1. The tilde over a particle symbol (e.g., $\tilde{g}$) means that it is the supersymmetric partner of a particle whose spin is less by ½ than the spin of the SM partner particle (e.g., a gaugino has a spin ½ while the spin of a sparticle is zero, but both are represented by the same overhead tilde sign).

The general diagrams and the details of the gauge interactions shown in Table 3 may be seen, for example in section 6 of "A supersymmetry primer" by S. Martin [35]. Further, in our work we have used only the gauge interactions and no Yukawa interactions which are known to have negligible contribution for the first-generation particles (as in our case) and phenomenologically also not required here (ref. same as above).

Now, we live in a universe which is dominated by ordinary matter and dark matter only. So, if they are supersymmetric, then as we shall see, dark matter particles must be the anti-superpartner of the ordinary matter particles and they must have been produced at the same time, i.e., in the cosmic epoch of inflation when they were in the plasma state. We further suppose that the matter particles and their anti-superpartners are the decay products of gauginos which are again the superpartners of the gauge bosons, carrier of all forces by their decay into and recombination of virtual particles and antiparticles as listed in Table 3.

It is now necessary to discuss some aspects of the decay products shown in Table 3. As regards any gauge boson, it decays into a virtual particle-antiparticle pair which are not stable and recombine to have the same boson again. Gauginos also decay in the same manner, but their decay products are of different types, one being an ordinary (natural) particle of spin half and the other an anti-scalar of spin zero. So, it is obvious that their decay period will be shorter and recombination period longer than those of spin half particle-antiparticle pair of the gauge boson. Though inflation caused a very rapid expansion of the universe, still it is possible that both virtual particle and its antiparticle of a gauge boson (supposing their decay and recombination time very short) remained within the limit of their attracting distance and so recombined again, whereas this did not happen for the decay products of a gaugino due to reasons stated above and so they became permanently separated from one another forming stable first generation ordinary and dark matter particles of the universe.

There is another aspect of the decay products of gauginos. We know that a supersymmetric partner of a chiral particle will also be chiral and of same mass, same charge, and other quantum numbers as of the particle itself except difference in their spins by 1/2. So, if a gaugino decays into a particle and an anti-supersymmetric particle, then this anti-supersymmetric particle must be the anti-supersymmetric partner of the same particle. This condition is, of course, satisfied for a neutral guagino (or boson) decay, but not for a charged one (see Table 3).

However, this problem is solved if we combine the decay products of a charged or neutral gaugino with the decay products of an oppositely charged or neutral boson of the same group and rearrange them to obtain a pair of a particle and its own anti-supersymmetric partner along with a pair of decay products which promptly recombine into a virtual boson again.

## 3 Combination Equations of the Vector Supermultipletes

Based on the above discussion, we now write down below all allowed combination equations in terms of their respective decay products as shown in Table 3. Considering the different types of particles and their interacting forces, we divide these equations into the following four sub-sections (sectors):

### 3.1 Equations for the Visible Baryons of the Strong Force Sector

Let us first consider the decay products ($u$, $\tilde{u}^*$, $u^\dagger$) of $\tilde{g}$ and $g$, as given in the Table 3 and write down all their possible combinations as given by the four Eqns. (1 – 4) below. However, the two up and down flavors of each $\tilde{g}$ and $g$ are subject to a $\mathrm{SU}(2)$ and a $\mathrm{U}(1)$ symmetry which subdivides these four decay equations into a set of three equations for $\mathrm{SU}(2)$ and only one equation for the singlets of them. Since, all these four equations are equivalent to one another, we can select any one of them as the singlet equation of $\tilde{g}$ and $g$. So, let us make a choice of Eqn. (1) for this singlet, and the other three Eqns. (2 – 4) for the triplets of $\tilde{g}$ and $g$. The specialty of this choice is that if we add the charges of only the up and down quarks, or antiquarks, or anti-squarks separately of these three equations, then in each case the sum of their respective charges will be 0.

Now each triplet of $\tilde{g}$ and $g$ chosen here are colored. So, each of them belongs to a color-octet group of $\mathrm{SU}(3)$ and a color-singlet $\mathrm{U}(1)$. Let the decay equations (5 - 7) for ($\tilde{g}_c$, $g_c$) represent the three color-octet decay equations of $\tilde{g}$ and $g$. Then Eqns. (2 – 4) will represent their corresponding three color-singlet equations.

In chromodynamics the color-singlet particles could be photons; but photon is massless and interacts with EM force. So, this choice is controversial and there are other theories too [49, see Ch. 8, pp. 285]. However, it is well known that electroweak $Z^o$ boson interacts with strong force particles and, accordingly, we get two more singlet equations being Eqns. Nos. (8 -9) given below for $Z^o$and $\widetilde{Z^o}$ as they are not colored.

$$\tilde{g} + g \rightarrow (u^{+2/3} + \tilde{u}^{*-2/3}) + (u^{+2/3} + u^{\dagger-2/3}) \rightarrow u^{+2/3} + \tilde{u}^{*-2/3} + g \quad (1)$$

$$\tilde{g} + g \rightarrow (u^{+2/3} + \tilde{u}^{*-2/3}) + (d^{-1/3} + d^{\dagger 1/3}) \rightarrow u^{+2/3} + \tilde{u}^{*-2/3} + g \quad (2)$$

$$\tilde{g} + g \rightarrow (d^{-1/3} + \tilde{d}^{*+1/3}) + (u^{+2/3} + u^{\dagger-2/3}) \rightarrow d^{-1/3} + \tilde{d}^{*+1/3} + g \quad (3)$$

$$\tilde{g} + g \rightarrow (d^{-1/3} + \tilde{d}^{*+1/3}) + (d^{-1/3} + d^{\dagger 1/3}) \rightarrow d^{-1/3} + \tilde{d}^{*+1/3} + g \quad (4)$$

$$\tilde{g}_c + g_c \rightarrow (u^{+2/3} + \tilde{u}^{*-2/3}) + (d^{-1/3} + d^{\dagger 1/3}) \rightarrow u^{+2/3} + \tilde{u}^{*-2/3} + g_c \quad (5)$$

$$\tilde{g}_c + g_c \rightarrow (d^{-1/3} + \tilde{d}^{*+1/3}) + (u^{+2/3} + u^{\dagger-2/3}) \rightarrow d^{-1/3} + \tilde{d}^{*+1/3} + g_c \quad (6)$$

$$\tilde{g}_c + g_c \rightarrow (d^{-1/3} + \tilde{d}^{*+1/3}) + (d^{-1/3} + d^{\dagger 1/3}) \rightarrow d^{-1/3} + \tilde{d}^{*+1/3} + g_c \quad (7)$$

$$\widetilde{Z^o} + Z^o \rightarrow (u^{+2/3} + \tilde{u}^{*-2/3}) + (u^{+2/3} + u^{\dagger-2/3}) \rightarrow u^{+2/3} + \tilde{u}^{*-2/3} + Z^o \quad (8)$$

$$\widetilde{Z^o} + Z^o \rightarrow (d^{-1/3} + \tilde{d}^{*+1/3}) + (d^{-1/3} + d^{\dagger 1/3}) \rightarrow d^{-1/3} + \tilde{d}^{*+1/3} + Z^o \quad (9)$$

The total number of equations represented by the above 9 combination equations of quarks, squarks, virtual gluons and $Z^o$ is given by

$[\mathrm{SU}(3)\mathrm{X}\ \mathrm{SU}(2)]\ \tilde{g}_c$ x $[\mathrm{SU}(3)\mathrm{X}\ \mathrm{SU}(2)]\ g_c + (4+2)$ x $\mathrm{U}(1)$

= (8x3) x (8x3) + 6

= 576 + 6 = 582 … … (10)

If we now look at the right-hand part of each of above 9 equations, we see that it contains one quark, one anti-squark and a gluon or a $Z^o$. This means that the total number of up and down quarks (anti-squarks) in all these equations is also 582 as given by Eqn. (10) above. But quarks (anti-squarks) are not natural particles unless three (up and down) of them combine to form a nucleon, $N$ (anti-snucleon, $\widetilde{N^*}$, as may be supposed). So, the total number of nucleons (anti-snucleons) made up of 582 quarks (anti-squarks) would be 194.

Now, in this sector we are concerned about the mass of $N$ only and keep aside $\widetilde{N}$ for the calculation of its mass, till we discuss the dark sector. In this paper we suppose that supersymmetry is unbroken. So, except $Z^o$ and $\widetilde{Z^o}$ all gluons and gluinos of this sector are massless. If quarks are supposed to be massive, even then these masses would be negligible compared to the mass of $Z^o$ , experimentally which is found to be $m_z$ = 91.19 GeV [49, see, pp. xiii]. On the other hand, nucleons are found to be quite massive. So, it may look surprising how a nucleon would get this large mass when quarks are so light particles. Moreover, the Higgs mechanism does not appear to provide this mass. However, in chromodynamics there is a mass formula to measure the masses of the baryons. But, for this we are to assume initial masses for u and d quarks [49, see Ch. 5, pp. 191]. On the other hand, we can solve this problem, simply by supposing that the total mass of two $Z^o$ bosons provide mass to all of 194 nucleons of this sector. Hence, we get the mass $m_N$ of each nucleon as given by

$$m_N = \frac{2mz}{194} = \frac{2 \text{ x } 91.19}{194} = 0.9401\ GeV = 940.10\ MeV \qquad \ldots \qquad (11)$$

(for c = 1). This calculated mass of a nucleon without any pre-assumption compares well with the current experimental value for a proton mass of 938.272 *MeV* and a neutron mass of 939.565 *MeV* [49, see pp. xiv]. Of course, this is an unexpected achievement of this theory. However, to prove that this result is not simply an accidental coincidence we will have to wait for discussion of the subsequent sectors and verify by applying this same method to calculate masses of other particles.

### 3.2. Equations for the Visible Leptons of the EM Force Sector.

This is a pure EM force sector, and we consider the equation numbers (12 – 17) for the visible fermions which are due to combinations of $\tilde{B}^0$ and $\widetilde{W}^0$ with $B^o$ and $Z^o$ as follows:

$$\tilde{B}^0 + B^o \rightarrow (e^{-1} + \tilde{e}^{*+1}) + (e^{-1} + e^{\dagger+1}) \rightarrow e^{-1} + \tilde{e}^{*+1} + B^o \qquad (12)$$

$$\widetilde{W}^0 + B^0 \rightarrow (e^{-1} + \tilde{e}^{*+1}) + (e^{-1} + e^{\dagger+1}) \rightarrow e^{-1} + \tilde{e}^{*+1} + B^o \qquad (13)$$

$$\tilde{B}^0 + Z^o \rightarrow (e^{-1} + \tilde{e}^{*+1}) + (\nu^o + \nu^{\dagger 0}) \rightarrow e^{-1} + \tilde{e}^{*+1} + Z^o \qquad (14)$$

$$\tilde{B}^0 + Z^o \rightarrow (e^{-1} + \tilde{e}^{*+1}) + (e^{-1} + e^{\dagger+1}) \rightarrow e^{-1} + \tilde{e}^{*+1} + Z^o \qquad (15)$$

$$\widetilde{W}^0 + Z^o \rightarrow (e^{-1} + \tilde{e}^{*+1}) + (\nu^o + \nu^{\dagger 0}) \rightarrow e^{-1} + \tilde{e}^{*+1} + Z^o \qquad (16)$$

$$\widetilde{W}^0 + Z^o \rightarrow (e^{-1} + \tilde{e}^{*+1}) + (e^{-1} + e^{\dagger+1}) \rightarrow e^{-1} + \tilde{e}^{*+1} + Z^o \qquad (17)$$

Here we suppose that $B^o$ mixes with $W^o$ to be converted into $Z^o$ in accordance with the well-known $EW$ theory. However, unlike $B^0$, $\tilde{B}^0$ does not mixes with $\widetilde{W}^0$ to produce a $\tilde{Z}^0$ gaugino, and so it remains as the same $\widetilde{W}^0$ to decay only into a pair of $e^{-1}$, $\bar{e}^{+1}$ like a $\tilde{B}^0$ gaugino.

In contrary to quarks, leptons $(\nu, e)$ are colorless and each of them has zero or unit charge. Moreover, neutrinos are supposed to have no mass whereas the electron has quite a negligible mass. So, unlike nucleons, these particles do not depend on $Z^o$ for their mass and directly become free particles in nature.

### 3.3. Equations for the Dark Leptons and Dark Baryons of the Weak Force Sector.

So far, we have written down all the possible equations for the ordinary visible particles of the universe. However, we have more equations for the particles of the weak force sector, too. But we have only two types of matter, namely ordinary matter and dark matter, in the universe. Hence all the equations of the weak sector must be due to dark leptons and dark baryons and accordingly they are divided into the following two sub-sections:

#### 3.3.1. Equations for the Dark Leptons of the Weak Force Sector.

Here we consider only dark leptonic decays of $\widetilde{W}$, $\tilde{Z}^0$ and W, Z bosons and their all-combinations as given by the following 12 equations (Nos. 18-29) with the only constraint (as we considered for the visible lepton sector above) that $\tilde{Z}^0$ (unlike $\widetilde{W}^0$) can decay only into $(e^{-1} + \tilde{e}^{+1})$ mode but not in $(\nu^o + \tilde{\nu}^0)$ mode:

$$\widetilde{W}^+ + W^- \rightarrow (\nu^0 + \tilde{e}^{*+1}) + (e^{-1} + \nu^{\dagger 0}) \rightarrow e^{-1} + \tilde{e}^{*+1} + Z^o \quad (18)$$

$$\widetilde{W}^- + W^+ \rightarrow (e^{-1} + \tilde{\nu}^{*0}) + (\nu^0 + e^{\dagger +1}) \rightarrow \nu^0 + \tilde{\nu}^{*0} + W^o \quad (19)$$

$$\widetilde{W}^0 + W^+ \rightarrow (\nu^o + \tilde{\nu}^{*0}) + (\nu^o + e^{\dagger +1}) \rightarrow \nu^o + \tilde{\nu}^{*0} + W^+ \quad (20)$$

$$\widetilde{W}^0 + W^+ \rightarrow (e^{-1} + \tilde{e}^{*+1}) + (\nu^o + e^{\dagger +1}) \rightarrow e^{-1} + \tilde{e}^{*+1} + W^+ \quad (21)$$

$$\widetilde{W}^0 + W^- \rightarrow (\nu^o + \tilde{\nu}^{*0}) + (e^{-1} + \nu^{\dagger 0}) \rightarrow \nu^o + \tilde{\nu}^{*0} + W^- \quad (22)$$

$$\widetilde{W}^0 + W^- \rightarrow (e^{-1} + \tilde{e}^{*+1}) + (e^{-1} + \overline{\nu}^0) \rightarrow e^{-1} + \tilde{e}^{*+1} + W^- \quad (23)$$

$$\widetilde{W}^+ + W^0 \rightarrow (\nu^0 + \tilde{e}^{*+1}) + (e^{-1} + e^{\dagger +1}) \rightarrow e^{-1} + \tilde{e}^{*+1} + W^+ \quad (24)$$

$$\widetilde{W}^- + W^0 \rightarrow (e^{-1} + \tilde{\nu}^{*0}) + (\nu^o + \nu^{\dagger 0}) \rightarrow \nu^o + \tilde{\nu}^{*0} + W^- \quad (25)$$

$$\widetilde{W}^0 + W^0 \rightarrow (\nu^o + \tilde{\nu}^{*0}) + (\nu^o + \nu^{\dagger 0}) \rightarrow \nu^o + \tilde{\nu}^{*0} + W^0 \quad (26)$$

$$\widetilde{W}^0 + W^0 \rightarrow (\nu^o + \tilde{\nu}^{*0}) + (e^{-1} + e^{\dagger +1}) \rightarrow \nu^o + \tilde{\nu}^{*0} + W^0 \quad (27)$$

$$\tilde{Z}^0 + Z^0 \rightarrow (e^{-1} + \tilde{e}^{*+1}) + (\nu^o + \nu^{\dagger 0}) \rightarrow e^{-1} + \tilde{e}^{*+1} + Z^0 \quad (28)$$

$$\tilde{Z}^0 + Z^0 \rightarrow (e^{-1} + \tilde{e}^{*+1}) + (e^{-1} + e^{\dagger +1}) \rightarrow e^{-1} + \tilde{e}^{*+1} + Z^0 \quad (29)$$

### 3.3.2. Equations of the Dark Baryons of the Weak Force Sector.

Like leptonic decays, W bosons and winos also have baryonic decays into $u$, $d$ quarks and their partner anti-squarks to have the following twelve equations (Nos.30-41). It may be noted that we have not considered baryonic decays of $(\tilde{Z}^0 + Z^0)$ here as these decays produce observable particles already included in the observable baryon sector above.

$$\widetilde{W}^+ + W^- \rightarrow (u^{2/3} + \tilde{d}^{*1/3}) + (d^{-1/3} + u^{\dagger-2/3}) \rightarrow d^{-1/3} + \tilde{d}^{*1/3} + W^0 \quad (30)$$

$$\widetilde{W}^- + W^+ \rightarrow (d^{-1/3} + \tilde{u}^{*-2/3}) + (u^{2/3} + d^{\dagger1/3}) \rightarrow u^{2/3} + \tilde{u}^{*-2/3} + W^0 \quad (31)$$

$$\widetilde{W}^0 + W^+ \rightarrow (u^{2/3} + \tilde{u}^{*-2/3}) + (u^{2/3} + d^{\dagger1/3}) \rightarrow u^{2/3} + \tilde{u}^{*-2/3} + W^+ \quad (32)$$

$$\widetilde{W}^0 + W^+ \rightarrow (d^{-1/3} + \tilde{d}^{*1/3}) + (u^{2/3} + d^{\dagger1/3}) \rightarrow d^{-1/3} + \tilde{d}^{*1/3} + W^+ \quad (33)$$

$$\widetilde{W}^0 + W^- \rightarrow (u^{2/3} + \tilde{u}^{*-2/3}) + (d^{-1/3} + u^{\dagger-2/3}) \rightarrow u^{2/3} + \tilde{u}^{*-2/3} + W^- \quad (34)$$

$$\widetilde{W}^0 + W^- \rightarrow (d^{-1/3} + \tilde{d}^{*1/3}) + (d^{-1/3} + u^{\dagger-2/3}) \rightarrow d^{-1/3} + \tilde{d}^{*1/3} + W^- \quad (35)$$

$$\widetilde{W}^+ + W^0 \rightarrow (u^{2/3} + \tilde{d}^{*1/3}) + (d^{-1/3} + d^{\dagger1/3}) \rightarrow d^{-1/3} + \tilde{d}^{*1/3} + W^+ \quad (36)$$

$$\widetilde{W}^- + W^0 \rightarrow (d^{-1/3} + \tilde{u}^{*-2/3}) + (u^{+2/3} + u^{\dagger-2/3}) \rightarrow u^{+2/3} + \tilde{u}^{*-2/3} + W^- \quad (37)$$

$$\widetilde{W}^0 + W^0 \rightarrow \left(u^{+2/3} + \tilde{u}^{*-2/3}\right) + \left(u^{+2/3} + u^{\dagger-2/3}\right) \rightarrow u^{+2/3} + \tilde{u}^{*-2/3} + W^0 \quad (38)$$

$$\widetilde{W}^0 + W^0 \rightarrow \left(u^{+2/3} + \tilde{u}^{*-2/3}\right) + \left(d^{-1/3} + d^{\dagger1/3}\right) \rightarrow u^{+2/3} + \tilde{u}^{*-2/3} + W^0 \quad (39)$$

$$\widetilde{W}^0 + W^0 \rightarrow \left(d^{-1/3} + \tilde{d}^{*1/3}\right) + \left(u^{+2/3} + u^{\dagger-2/3}\right) \rightarrow d^{-1/3} + \tilde{d}^{*1/3} + W^0 \quad (40)$$

$$\widetilde{W}^0 + W^0 \rightarrow \left(d^{-1/3} + \tilde{d}^{*1/3}\right) + \left(d^{-1/3} + d^{\dagger1/3}\right) \rightarrow d^{-1/3} + \tilde{d}^{*1/3} + W^0 \quad (41)$$

### 3.4. Calculation of Mass Density Parameters.

We now want to calculate both the initial and final masses for all the equations of both the dark lepton and the dark baryon sectors.

For this, let us go back to the observable (or dark) baryon sector. Here we see that for every quark there is its anti-squark which must have the same mass but equal and opposite charge and baryon number. So, like quarks which produced natural nucleons (*N*), the partner anti-squarks should also produce similar anti-snucleons ($\widetilde{N^*}$) and their number should also be equal to the number of the natural nucleons. This would mean that the universe would have no asymmetry of charge, baryon number and particle-antiparticle number, provided we can calculate the natural mass ratio from the calculated mass of the baryons and the mass of the anti-snucleons. However, in Eqn. (11), we have already calculated the mass of the nucleon to be 940.10 *MeV,* which agrees with its natural mass.

So, the next question is: what is the mass of the anti-snucleon? For this we suppose that these anti-snucleons are the dark matter of the universe and they get their mass $M_{DM}$ from the winos and zenos $(\widetilde{W}, \tilde{Z})$ of the two dark sectors.

But how would we know the mass of a wino or zeno? Experimentally we know the masses of *W* and *Z* bosons to be $m_W$ = 80.42 *GeV* and $m_Z$ = 91.19 *GeV* [49 see pp. xiv]. But as regards gauginos, no experiment has yet been successful to measure their masses. However, we shall discuss possibilities of experimental success for this mass measurement in the last section of this paper.

Now theoretically we take the advantage of the mass, energy and momentum conservation laws, according to which if a particle A decays into a pair of particles B and its antiparticle $\overline{\mathrm{B}}$ of same mass, then in absence of kinetic or any other energies (as in vacuum polarization) the mass of B will be at best (called its threshold mass) equal to half the mass of A [49 see Ch. 3, pp.103] . Applying this principle and since in vacuum polarization a $W$ $(Z^o)$ boson can be supposed to decay into a pair of $\widetilde{W}$ $(\tilde{Z}^0)$ and its anti-particle $\widetilde{W}^*$ $(\tilde{Z}^{0*})$, we take the mass of $\widetilde{W}$ to be equal to half of the mass of *W* boson and the mass of $\tilde{Z}^0$ equal to half the mass of $Z^o$ boson. So, considering the above two dark sectors, the total mass of the winos and zenos is given by

$M_{DM}$ = masses of (10 $\widetilde{W}$ and 2 $\tilde{Z}$ of the dark lepton sector + 12$\widetilde{W}$ from dark baryon sector)

= 5 $m_w$ + $m_z$ + 6 $m_w$

= (91.19 + 11 x 80.42) *GeV* = 975.81 *GeV* (42)

where we have put $m_{\widetilde{W}}$ $(m_{\tilde{Z}})$ = ½ $m_w$ $(m_z)$ as deduced above. Therefore, the mass of each anti-snucleons (the dark nucleon) $(\widetilde{N^*})$ is given by

$m_{\widetilde{N^*}}$ = 975.81 *GeV*/194 = 5.004 *GeV* … … (43)

The ratio of this calculated mass of dark nucleon vs. the calculated mass of the ordinary nucleon (Eq. 13) is:

$m_{\widetilde{N}}/ m_N$ = 5.00415 / 0.940 = 5.324 … (44)

This means that the theoretical dark nucleon is 5.324 times heavier than the calculated nucleon mass. There is no direct experimental proof for this $\widetilde{N^*}$ mass, but from the (2018) precision result of Planck observations, we find that the ratio of dark matter density vs. baryonic matter density is

= $\Omega_c$ / $\Omega_b$ = (0.2646 ± 0.0003) / (0.0493 ± 0.0003) = 5.367 … (45)

which shows that our calculated result agrees with the Planck observational result [2] within an error range of 0.8 % only. This proves that none of our theoretical measurements of $m_N$ and $m_{\widetilde{N}}$ give accidental agreement with their respective practical values.

Based on the above derivations, we can now proceed to calculate the Planck results from the gauge supermultiplet equations given above (Eqn. Nos. 1 –9 and 12 – 41). In each of these equations, the sum of the masses of two vector supermultiplets on the left-hand side will be termed as the initial mass and the mass of the last boson on the right-hand side will be termed as the vacuum-polarization boson being responsible for the dark energy of the universe. Moreover, the masses of the chiral super particle-anti-sparticle pair on the right-hand side is supposed to be negligibly small. So, the sector wise total initial mass of the vector supermultiplets and the total mass of the vacuum-polarization bosons (being the total dark energy) for all the above equations of the four sectors are given by:

Sector 3.1. Visible baryons (Eqns. 1 – 9): As calculated above this sector has a total of 582 equations. But except 2 $\tilde{Z}^0$'s and 2 $Z^o$ all the gluinos ($\tilde{g}$) and gluons ($g$) of this sector are massless.

Therefore, the total initial mass of this sector is = 2 x (91.19 / 2 + 91.19) *GeV*

$m_{inl}$ = 273.57 *GeV* … (46)

and the mass of the two vacuum-polarization bosons ($Z^o$)

$m_{DE}$ = 2 x 91.19 = 182.38 *GeV* (47)

Sector 3.2. Visible leptons (Eqns. 12 – 17): Here there are 6 equations in which $B^o$ and $\tilde{B}^0$ are supposed to be massless.

Therefore, for this sector the total initial mass is

$m_{inl}$ = ($3\tilde{B}^0$ mass) + (3 $\tilde{W}^0$ mass) + ($2B^o$ mass) + ($4Z^o$ mass)

= (3 x 0 + 3/2 x 80.42 + 2 x 0 + 4 x 91.19) *GeV* = 485.39 *GeV,* (48)

and the total vacuum-polarization boson mass is

$m_{DE}$ = 2 $B^o$ mass + 4 $Z^o$ mass = 364.76 *GeV* (49)

Sector 3.3.1. Dark lepton sector, (Eqns. 18 – 29): Here there are 12 equations for which the total initial mass is

$m_{inl}$ = (10 $\tilde{W}$ mass) + (10 W mass) + (2 $\tilde{Z}$ mass) + (2 Z mass)

= (15 x 80.42 +3 x 91.19) *GeV*

= 1479.87 *GeV* … (50)

and the total vacuum-polarization boson mass = 9 W mass + 3 Z mass

$m_{DE}$ = (9 x 80.42 + 3 x 91.19) *GeV* = 997.35 *GeV* (51)

Sector 3.3.2: In this Dark Baryon sector, (Eqns. 30 – 41) there are 12 equations of total initial mass

$m_{inl}$ = (12 $\tilde{W}$ mass) + (12 W mass) 18 x 80.42 *GeV* = 1447.56 *GeV* (52)

and the total vacuum-polarization boson mass

$m_{DE}$ = 12 W mass = 965.04 *GeV* (53)

Hence the total initial mass of all these four sectors is given by

$M_{total}$ = (273.57 + 485.39 + 1479.87 + 1447.56) *GeV*

= 3686.39 *GeV* … (54)

and the total vacuum-polarization (dark energy) boson mass of all these four sectors is

$$M_{DE} = (182.38 + 364.76 + 997.35 + 965.04)\ GeV = 2509.53\ GeV \qquad (55)$$

Now we calculate the fractional densities of different sector particles and the dark energy bosons as follows:

Visible baryon density: In section 2.1, the total mass of all the baryons is given by $M_N = 2m_z$ =182.38 *GeV* (Eqn.11). So, dividing this mass by the total initial mass of all the sectors (Eqn.54), we get the fractional baryonic matter density for all these sectors:

$$D_N = \frac{M_N}{M_{total}} = \frac{182.38}{3686.39} = 0.04947 \quad \ldots \qquad (56)$$

Next, we calculate the fractional dark matter density for our model. According to Eq. (42), the total dark matter mass as calculated from the equations of the dark sectors is given by

$$M_{DM} = 975.81\ GeV$$

Hence the fractional dark matter density is

$$D_{DM} = \frac{M_{DM}}{M_{total}} = \frac{975.81}{3686.39} = 0.2647 \qquad (57)$$

Now, if we combine the above two densities, we get the total density of matter as given by

$$D_{matter} = 0.04947 + 0.2647 = 0.3142 \quad \ldots \qquad (58)$$

Now, let us calculate the mass density for the vacuum-polarization boson mass. Equation (55) gives the total mass of all these bosons of the four sectors as

$$M_{DE} = 2509.53\ GeV$$

Therefore, the fractional density of these bosons would be

$$D_{DE} = \frac{M_{DE}}{M_{total}} = \frac{2509.53}{3686.39} = 0.68076 \quad \ldots \qquad (59)$$

The calculated results obtained above are summarized in Table 4 below for convenience and comparison of them with the respective Planck results as listed in the last column of Table 4. It shows that our results are in good agreement with those of Planck results.

TABLE 4

| Sector | Total Mass ($GeV$) | Calculated Density (D) $= \frac{sector\ total\ mass}{total\ initial\ mass}$ | Planck (2018) Result [2] |
|---|---|---|---|
| Initial mass | $M_{total}$ = 3686.39 | $\frac{3686.39}{3686.39}$ = 1 | |
| Visible Baryon mass | $M_N$ = 182.38 | $\frac{182.38}{3686.39}$ = 0.04947 | $\Omega_b$ = 0.0493 ± 0.0003 |
| Dark matter mass | $M_{DM}$ = 975.81 | $\frac{975.81}{3686.39}$ = 0.2647 | $\Omega_c$ = 0.2646 ± 0.0003 |
| Total matter mass | $M_{matter}$ = 1158.19 | $\frac{1158.19}{3686.39}$ = 0.3142 | $\Omega_m$ = 0.3153 ± 0.0073 |
| vacuum-polarization boson mass | $M_{DE}$ = 2509.53 | $\frac{2509.53}{3686.39}$ = 0.68076 | $\Omega_\Lambda$ = 0.6847 ± 0.0073 |

Table 4. Summary of our calculated densities and comparison of them with the Planck Collaboration (2018) results of precision (68 %, Planck TT, TE, EE + lowE + lensing) measurement of the densities of baryonic matter ($\Omega_b$), the dark matter ($\Omega_c$), total matter ($\Omega_m$) and dark energy ($\Omega_\Lambda$) [2]

I shall not proceed further to calculate other Planck parameters e.g., $\mathrm{H}_0$, $\sigma$ etc. as they are mostly related to the expansion and acceleration of the universe not being considered here. So, I now proceed to the next section, but before that I would like to show that the Higgs boson mass is closely related to all the equations of the natural boson Multiplets shown above. These relationships can be deduced in three ways as follows:

(1) Relationship with the dark energy Bosons:

According to Eq. (55) the total mass of all the DE bosons is 2509.53 $GeV$. This mass corresponds to 32 equations of all the natural bosons (massive EW bosons and massless $B^0$ boson) of which 20 equations are for neutral bosons and other 12 are for W $\pm$ charged bosons. Now suppose that the mass of these 12 bosons combines with the mass of 20 neutral bosons to produce 20 neutral particles. Then the mass of each of these 20 particles would be

$$M_h = 2509.53 / 20 = 125.4765\ GeV \tag{60}$$

which is the mass of a Higgs boson with an error of order 0.293 % only compared to its experimental value of 125.11 $GeV$. This error may be compared with the SM estimate for the mass of the Higgs boson (before its discovery) which had a range of roughly 90 to 135 $GeV$.

(2) Relationship with the initial gauginos and gauge bosons of all the equations:

Since the total initial mass of all the equations described here are due to only massive gauginos and gauge bosons of their respective equations, it is obvious that they get this mass only from the Higgs boson. But the mass distribution in these equations is not uniform. So, we make the following counting rules for the Higgs boson: (i) if both the gaugino and the gauge boson of an equation are massive, then we shall take the equation has got on the average the mass of one Higgs boson. However, (ii) if in an equation either of the gaugino and the gauge boson is massive, but the other is massless, then the equation has got only ½ of the Higgs mass, and finally, (iii) if both these quantities are massless, then the equation has got no mass from the Higgs boson. Applying these conditions, we have a total of 28 full-weightage equations (2 from each of ordinary baryon and lepton sectors and 24 from dark sectors), and 3 half-weightage equations from the ordinary lepton sector. All other equations including that of $\tilde{B}^0$ and $B^o$ are of zero-weightage and not considered. Thus, the total effective number of Higgs bosons in these equations is = 28 + 3x1/2 = 29.5, and for them the total initial mass is

$M_{total}$ = $3686.39 GeV$ { as shown by Eq. No. (54)}.

Hence, the mass of the a Higgs boson is $M_h = \frac{M_{total}}{29.5} = \frac{3686.39}{29.5}$ = 124.962 $GeV$ (61)

Which shows a less error (0.118%) for $M_h$ than that of Eq. (60).

(3) Relationship with the gauginos and the masses of the ordinary and dark nucleons:

In the ordinary sectors of baryons and leptons there are a total of 5 winos and zenos, whereas in the two dark sectors the total number of them is 24. Let us suppose that each wino (zeno) of the ordinary sectors is related to the mass of an ordinary nucleon (N) and each wino (zeno) of the dark sectors is related to the mass of dark nucleon ($\widetilde{N^*}$). If the masses of all these nucleons combine to constitute the mass of a Higgs boson, this mass would be

$M_h$ =5 x $m_N$ + 24 x $m_{\widetilde{N^*}}$ = 5 x 0.9401 + 24 x 5.004 = 124.7964 (62)

Which has an error of 0.25% with respect to the experimental mass.

## 4. Dark Baryon and Solution to Matter-Antimatter Asymmetry and Baryon Number and Lepton Number Violation.

The mass of an anti-snucleon (dark nucleon $\tilde{N}^*$) has already been calculated to have a value $m_{\tilde{N}}$ = 5.004 $GeV$, (Eq. 43).

Let us now look for its composition and other properties. We get these properties from the quark anti-superpartners of the gluino-gluon decay equations shown in sector 3.1. Here we find that for each up quark ($u$) and down quark ($d$), there are their corresponding anti-superpartners : anti-up squark ($\tilde{u}^*$) and anti-down squark ( $\tilde{d}^*$). So, like ordinary nucleons these anti-squarks also form anti-snucleons, one is negatively charged, anti-sproton $\tilde{p}^*(\tilde{u}^*\tilde{u}^*\tilde{d}^*)^{-1}$ and the other is neutral, anti-sneutron, $\tilde{n}^*(\tilde{u}^*\tilde{d}^*\tilde{d}^*)^0$. Moreover, the baryon number of each anti-snucleon is equal and negative to the baryon number of an ordinary nucleon. They are also produced in equal numbers. Hence, both the total charge and baryon numbers of the universe is zero, which means that there is no asymmetry for these particles and anti-particles in the universe.

Similar arguments apply for the leptons ($\nu^0,\ e^{-1}$) and their anti-spartners ($\tilde{\nu}^{*0}$, $\tilde{e}^{*+1}$), too, with the difference that they are not composite but fundamental natural particles and anti-sparticles having opposite integral charges and lepton numbers instead of baryon numbers.

Now the question is, what is the real composition of the dark matter of the universe? Since the ordinary matter of the universe is neutral but has both baryon number and lepton number, the dark matter must also be neutral, and it must have the same but opposite baryon number and lepton number. As discussed above, each of these numbers is universally conserved. But the reality is that the dark matter halos stay too far away from the galactic ordinary matter. So, if the dark matter halos are made of anti-sprotons and anti-sneutrons, all the anti-selectrons should remain in the same halo. This can make dark matter of two varieties.

The first variety can happen if the anti-sproton is unstable and it is slightly heavier than the mass of the stable anti-sneutron. In that case the anti-sproton would decay into an anti-sneutron, an anti-selectron and an anti-sneutrino:

$$\tilde{p}^{*-1} \quad \rightarrow \quad \tilde{n}^{*} + \tilde{e}^{*-1} + \tilde{\nu}^{*0} \qquad \ldots \qquad (63)$$

In this case $\tilde{e}^{*-1}$ would annihilate the free $\tilde{e}^{*+1}$ and we get the dark matter halo entirely made of only $\tilde{n}^{*}$ and $\tilde{\nu}^{*0}$ only.

The second variety would be obtained if the interacting anti-sparticles evolve like the BB nucleosynthesis [8] process of the SM and according to the GSW theory of weak interactions [ 50, Ch. 20, pp. 707]. In this case $\tilde{p}^{*}$, $\tilde{n}^{*}$ and $\tilde{e}^{*}$ would form bound states of different anti-satoms to make a heavy dark matter halo.

## 5. Conclusion.

In conclusion it can be said that the MSSM theory has been successful in explaining the Planck (2018) cosmological parameter precision values and thereby proves the existence of dark matter and the validity of the Lambda-CDM model. It further establishes that the whole mass and energy of the universe is entirely due to the mass and energy of the weak bosons and the weak gauginos and in that sense dark matter is very much real and not an exotic substance. Additionally, it has been possible to calculate the masses of an ordinary nucleon, the dark matter nucleon, and the Higgs boson based on the theoretically determined masses of the weak gauginos and the known masses of the weak bosons. Dark matter being made up of either anti-sneutrons or bound anti-satoms has zero charge and, it solves the matter-antimatter asymmetry problem of the universe. Further, the baryon and lepton numbers of them are opposite to those of ordinary baryons and leptons, and therefore, both these numbers are conserved. Finaly, we have seen that Higgs boson is intimately related to all of gauge bosons, gauginos, nucleons and even anti-snuclens. So, it is hoped that the theoretically determined masses of the wino and zeno being equal to half of the masses of W and Z bosons respectively can be confirmed experimentally by measuring the missing energy of Higgs boson decay products in the high energy pp collisions in the LHC. This is important because the whole theory described above is dependent on these masses in addition to the known masses of the weak bosons.